# DESIGN OF THE LBNF BEAMLINE *

V. Papadimitriou[†], K. Ammigan, J. Anderson Jr., K. E. Anderson, R. Andrews, V. Bocean, C. F. Crowley, N. Eddy, B. D. Hartsell, S. Hays, P. Hurh, J. Hylen, J. A. Johnstone, P. Kasper, T. Kobilarcik, G. E. Krafczyk, B. Lundberg, A. Marchionni, N. V. Mokhov, C. D. Moore, D. Pushka, I. Rakhno, S. D. Reitzner, P. Schlabach, V. Sidorov, A. M. Stefanik, S. Tariq, L. Valerio, K. Vaziri, G. Velev, G. Vogel, K. Williams, R. M. Zwaska, Fermilab, Batavia, IL 60510, USA
C. Densham, STFC/RAL, Didcot, Oxfordshire, OX11 0QX, UK

*Abstract*

The Long Baseline Neutrino Facility (LBNF) will utilize a beamline located at Fermilab to provide and aim a neutrino beam of sufficient intensity and appropriate energy range toward the Deep Underground Neutrino Experiment (DUNE) detectors, placed deep underground at the SURF Facility in Lead, South Dakota. The primary proton beam (60-120 GeV) will be extracted from the MI-10 section of Fermilab's Main Injector. Neutrinos will be produced when the protons interact with a solid target to produce mesons which will be subsequently focused by magnetic horns into a 194m long decay pipe where they decay into muons and neutrinos. The parameters of the facility were determined taking into account the physics goals, spatial and radiological constraints, and the experience gained by operating the NuMI facility at Fermilab. The Beamline facility is designed for initial operation at a proton-beam power of 1.2 MW, with the capability to support an upgrade to 2.4 MW. LBNF/DUNE obtained CD-1 approval in November 2015. We discuss here the design status and the associated challenges as well as the R&D and plans for improvements before baselining the facility.

## INTRODUCTION

The Beamline is a central component of LBNF and it is expected to produce the highest power neutrino beam in the world. Its driving physics consideration is the study of long baseline neutrino oscillations. This study will take place through the Deep Underground Neutrino Experiment which has detectors located both at the Fermilab site and at the Sanford Underground Research Facility (SURF), about 1300 km away. LBNF/DUNE achieved the CD-1 conceptual design milestone in November 2015.

The beamline facility is expected to be fully contained within Fermilab property. The primary proton beam, in the energy range of 60-120 GeV, will be extracted from the Main Injector's (MI) [1] MI-10 section using "single-turn" extraction. For 60 - 120 GeV operation and with the MI upgrades implemented for the NOvA experiment [2] as well as with the expected implementation of the accelerator Proton Improvement Plan, phase II (PIP-II) [3], the fast, single turn extraction will deliver 7.5 x $10^{13}$ protons in one MI machine cycle (0.7 sec - 1.2 sec) to the LBNF target in 10 µs. The beam power is expected to be from 1.03 to 1.20 MW in the energy range of 60 to 120 GeV [3]. The charged mesons produced by the interaction of the protons with the target are sign selected and focused by two magnetic horns into the decay pipe towards the far detector. These mesons are short-lived and decay into muons and neutrinos. At the end of the decay region, an absorber is needed to remove the residual hadrons remaining at the end of the decay pipe. The neutrino beam is aimed 4850 ft underground at SURF in South Dakota.

A wide band neutrino beam is needed to cover the first and second neutrino oscillation maxima, which for a 1300 km baseline are expected to be approximately at 2.4 and 0.8 GeV. The beam must provide a high neutrino flux at the energies bounded by the oscillation peaks and thus we are optimizing the beamline design for neutrino energies between 0.5 and 5 GeV. The initial operation of the facility will be at a beam power of 1.2 MW on the production target, however some of the initial implementation will have to be done in such a manner that operation at 2.4 MW can be achieved without major retrofitting. Such a higher beam power is expected to become available in the future with additional improvements in the Fermilab accelerator complex [4]. In general, components of the LBNF beamline system which cannot be replaced or easily modified after substantial irradiation at 1.2 MW operation are being designed for 2.4 MW. Examples of such components are the hadron absorber and the shielding of the target chase and decay pipe.

The LBNF Beamline design has to implement as well a stringent radiological protection program for the environment, workers and members of the public. The relevant radiological concerns: prompt dose, residual dose, air activation and water activation have been extensively modelled and the results are incorporated in the system design. A most important aspect of modelling at the present design stage is the determination of the necessary shielding thickness and composition in order to protect the ground water and the public and to control air emissions.

This paper is a snapshot of the present status of the design, detailed in the 2015 Conceptual Design Report [5, 6].

## STATUS OF THE DESIGN

Figure 1 shows a longitudinal section of the LBNF beamline facility. At MI-10 there is no existing extraction enclosure and we are minimizing the impact on the MI by introducing a 15.6 m long beam carrier pipe to transport the beam through the MI tunnel wall into the new LBNF enclosure. The extraction and transport components send the proton beam over a man-made embankment/hill whose apex is at 18.3 m above ground and with a footprint of

___________________________________________
* Work supported by DOE, contract No. DE-AC02-07CH11359
† vaia@fnal.gov

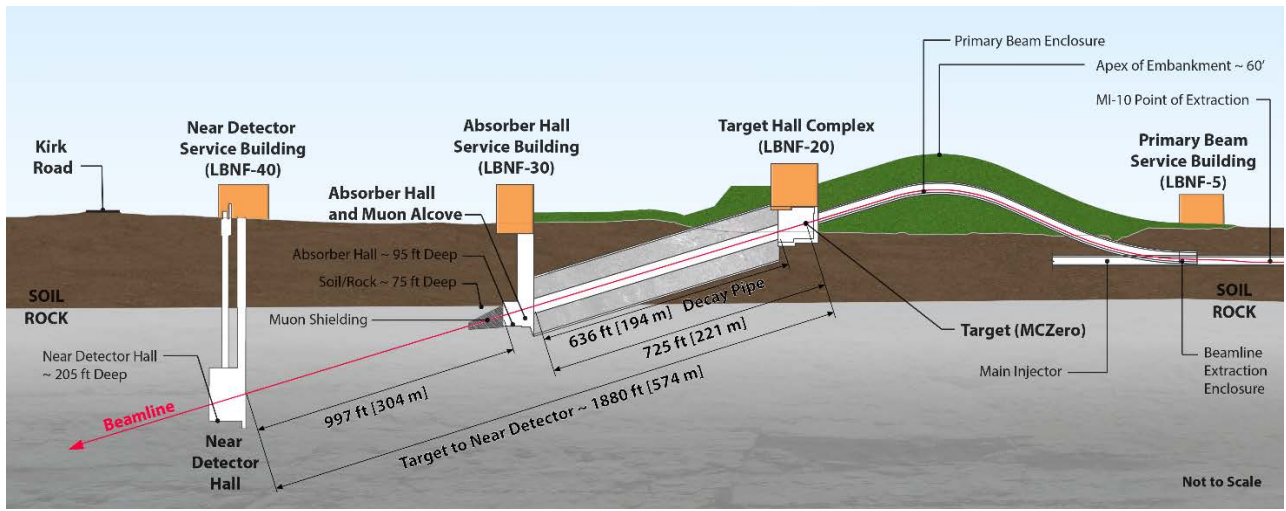

Figure 1: Longitudinal section of the LBNF beamline facility at Fermilab. The beam comes from the right, the protons being extracted from the MI-10 straight section of the MI.

~21,370 m$^2$. At the top of the hill the beam then will be bent downward toward a target located at grade level. The overall bend of the proton beam is 7.2º westward and 5.8º downward to establish the final trajectory toward the far detector.

In this shallow beamline design, because of the presence of a local aquifer at and near the top of the rock surface, an engineered geomembrane barrier and drainage system between the shielding and the environment prevents the contamination of groundwater from radionuclides. The decay pipe shielding thickness has been determined to be 5.6 m of concrete (see Fig. 2).

*Beamline Scope*

The LBNF Beamline scope includes a primary (proton) beamline, a neutrino beamline and associated conventional facilities. The primary beamline elements necessary for extraction and transport include kicker, Lambertson, C, dipole, quadrupole and corrector magnets connected by vacuum pipes and beam monitoring equipment (Beam-Position Monitors, Beam-Loss Monitors, Beam-Profile Monitors, and Beam-Intensity monitors). The magnets (79 in total) are conventional, MI design, and the magnet power supplies are a mixture of new, MI design power supplies and refurbished Tevatron power supplies. The beam optics accommodates a range of spot sizes on the target (1-4 mm RMS) in the energy range of interest and for beam power up to 2.4 MW, and the beam transport is expected to take place with negligible losses.

The neutrino beamline closely follows the design of the NuMI focusing system. It includes in order of placement (1) a beryllium window that seals off and separates the evacuated primary beamline from the neutrino beamline, (2) a baffle collimator assembly to protect the target and the horns from mis-steered beam, (3) a target, (4) two magnetic horns. The LBNF horns operate at higher current and lower pulse width compared to NuMI. These elements are all located inside a heavily shielded, air-filled, air/water-cooled vault, called the target chase (see Fig. 2), that is isolated from the decay pipe at its downstream end by a replaceable, thin, metallic window. The target chase has sufficient length and cross-section to accommodate an optimized focusing system. In order to mitigate potential corrosion issues and release of air-born radionuclides we are studying alternative gases (nitrogen and helium) for the atmosphere of the target chase. A 194 m long, 4m in diameter helium-filled, air-cooled decay pipe follows the chase and within it pions and kaons decay to neutrinos. At its end is the hadron absorber which must contain the energy of the particles that exit the decay pipe. The absorber core consists of replaceable aluminium and steel water-cooled blocks. Outside of the core we have steel and concrete shielding that is cooled by forced-air. Approximately 40% of the beam power is deposited in the target chase and surrounding shielding, 30% in the decay pipe and 30% in the absorber.

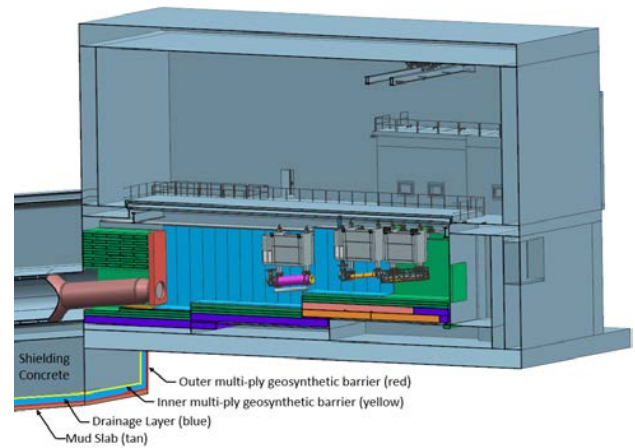

Figure 2: Schematic of the upstream portion of the LBNF neutrino beamline showing the major components of the neutrino beam. The steel shielding above the beam components and chase is not shown for clarity. Inside the chase, from right to left, one can see the horn-protection baffle and the target mounted on a carrier, and the two focusing horns. The decay pipe follows.

Radiation damage, cooling of elements, radionuclide mitigation, remote handling and storage of radioactive components are essential considerations for the conceptual design of the neutrino beamline.

The reference design for the LBNF target and focusing horns is an upgraded version of the NuMI low-energy target/horn system that was used for ~8 years to deliver neutrino beam to the NuMI experiments. The target is 95 cm long, two interaction length, with graphite fins and is water cooled. It extends 50 cm into the first horn. The horns' inner conductors are a double parabolic shape and can be operated at 230 kA (NuMI horns were originally designed to operate up to 200 kA). A new horn power supply will be required with current pulse width of 0.8 ms.

*Ongoing neutrino beam optimization studies*

There is ongoing effort to increase the physics reach of DUNE by optimizing the target and horn designs for maximal CP violation sensitivity. While considering two-horn configurations, the optimization led to extremely long horns which would be challenging to build. Figure 3 shows a three-horn configuration (~ 3m long horns) where the horn inner conductors are combinations of cylindrical and conical sections and the target is inserted entirely into the first horn's inner conductor. The targets considered in the optimization studies are up to 2m long, with graphite or beryllium fins, beryllium spheres and a graphite cylinder.

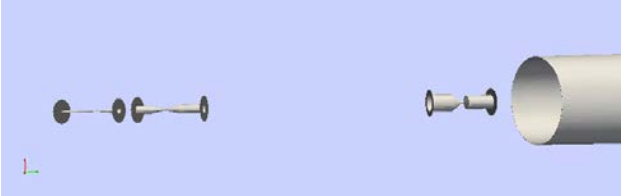

Figure 3: An optimized three-horn configuration. The proton beam comes from the left and the mesons produced at the target enter the decay pipe shown on the right.

*Physics reach with the reference design and alternative options*

Figure 4 shows the comparison of un-oscillated $\nu_\mu$ fluxes for the reference design and two other studied modifications. One of them is for two optimized horns and the other for three optimized horns. Both optimized designs use a NuMI-style carbon fin target that was 2.5 m long for the two-horn optimization and 2.0 m long for the three-horn optimization. The reference design is for 120 GeV protons while the optimized designs are for 80 GeV (two-horns) and 62 GeV (three-horns). The CP violation sensitivity does not change much though for proton energies in the range of 60 and 120 GeV. We note that with the optimized configurations the neutrino flux is increased between the first and second oscillation maximum.

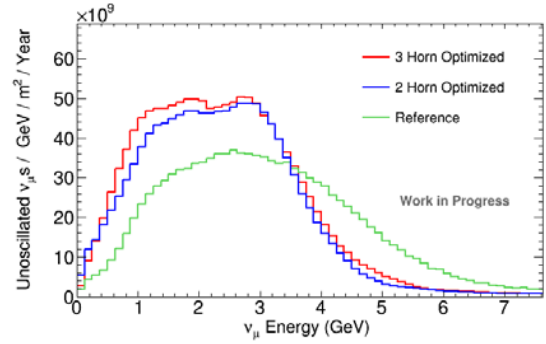

Figure 4: Neutrino mode fluxes of muon neutrinos as a function of neutrino energy for the reference design and a 2-horn and 3-horn optimized design.

## CONCLUSION

We described above the conceptual design of the LBNF beamline. We are now in the process of advancing this design and are exploring some alternative options for the target and horn designs that could enhance the physics capabilities of DUNE. We will complete the evaluation of these alternatives before baselining.

## ACKNOWLEDGMENT

We would like to thank the broader LBNF beamline team for their numerous contributions towards developing and costing the beamline conceptual design, and advancing it in all areas towards baselining. We also thank the DUNE collaboration for contributions to the beamline optimization and detailed studies of the neutrino fluxes.